\documentclass[superscriptaddress,showpacs,twocolumn,amsmath,amssymb,prl]{revtex4} 

\usepackage{graphicx} 

\begin{document}


\title{Irreversibility on the Level of Single-Electron Tunneling}


\author{B.~K\"ung}
\email{kuengb@phys.ethz.ch}
\affiliation
{Solid State Physics Laboratory, ETH Zurich,
8093 Zurich, Switzerland}

\author{C.~R\"ossler}
\affiliation
{Solid State Physics Laboratory, ETH Zurich,
8093 Zurich, Switzerland}

\author{M.~Beck}
\affiliation
{Institute for Quantum Electronics, ETH Zurich,
8093 Zurich, Switzerland}

\author{M.~Marthaler}
\affiliation
{Institut f\"ur Theoretische Festk\"orperphysik
and DFG Center for Functional Nanostructures (CFN),
Karlsruhe Institute of Technology, 76128 Karlsruhe, Germany}

\author{D.~S.~Golubev}
\affiliation
{Institut f\"ur Nanotechnologie, Karlsruhe Institute of Technology, 76021 Karlsruhe, Germany}

\author{Y.~Utsumi}
\affiliation
{Department of Physics Engineering, Faculty of Engineering, Mie University, Tsu, Mie, 514-8507, Japan}

\author{T.~Ihn}
\affiliation
{Solid State Physics Laboratory, ETH Zurich,
8093 Zurich, Switzerland}

\author{K.~Ensslin}
\affiliation
{Solid State Physics Laboratory, ETH Zurich,
8093 Zurich, Switzerland}

\date{July 21st, 2011}

\pacs{05.40.-a, 05.70.Ln, 73.63.Kv, 73.23.Hk}

\begin{abstract}
We present a low-temperature experimental test of the fluctuation theorem for electron transport through a double quantum dot. The rare entropy-consuming system trajectories are detected in the form of single charges flowing against the source-drain bias by using time-resolved charge detection with a quantum point contact. We find that these trajectories appear with a frequency that agrees with the theoretical predictions even under strong nonequilibrium conditions, when the finite bandwidth of the charge detection is taken into account.
\end{abstract}

\maketitle

The second law of thermodynamics states that a macroscopic system out of thermal equilibrium will irreversibly move toward equilibrium driven by a steady increase of its entropy. This macroscopic irreversibility occurs despite the time-reversal symmetry of the underlying equations of motion. Also a microscopic system will undergo an irreversible evolution on a long time scale, but over a sufficiently short observation time $\tau$, both entropy-producing trajectories as well as their their time-reversed entropy-consuming counterparts occur. It is only because of the statistics of these occurrences that a long-term irreversible evolution is established. This phenomenon is described by the fluctuation theorem \cite{Bochkov77,Evans93}.

Irrespective of the description of the trajectories being system-specific, the fluctuation theorem (FT) relates the probabilities $P_\tau(\Delta S)$ for processes that change the entropy of the system by an amount $\Delta S$ during an arbitrary time $\tau$ by the equation
\begin{equation}
\label{eq:FlucThm_FT}
\frac{P_{\tau}(\Delta S)}{P_{\tau}(-\Delta S)} = e^{\Delta S/k_B},
\end{equation}
where $k_B$ is the Boltzmann constant. In a seminal work, Wang {\it et al.}~\cite{Wang02} tested the FT by measuring force trajectories of a micron-scale latex bead in a liquid. Since then, the FT has been tested in other systems \cite{Garnier05,Collin05,Schuler05}, but all of those earlier experiments were carried out in a classical regime at room temperature. Experiments in the quantum regime \cite{Kurchan00,Tasaki00,Esposito09} have long been anticipated, and the use of quantum-coherent mesoscopic conductors may lead to this goal. However, at their typical operation temperatures $T$ below $1 \, \mathrm{K}$, the requirement to resolve tiny fluctuations on the energy scale $k_BT$ becomes an increasingly challenging task. An interesting recent result in this direction has been the verification of exact relations between current and current noise as functions of the source-drain voltage across an Aharonov--Bohm interferometer \cite{Nakamura10}. These relations are naturally derived from the FT [Eq.~\eqref{eq:FlucThm_FT}] as well as Onsager--Casimir and fluctuation-dissipation relations \cite{Foerster08,Saito08}.

As a step toward the direct test of Eq.~\eqref{eq:FlucThm_FT} in the quantum regime, we verify the fluctuation theorem in single-electron tunneling \cite{Tobiska05,Andrieux06,Esposito07} at low temperatures, although our experiment is carried out in the regime of classical charge counting. We employ real-time detection of single-electron charging \cite{Vandersypen04,Schleser04} in quantum dots (QDs). Monitoring the charge state of two QDs coupled both in series and to source and drain electrodes (double quantum dot [DQD]) allows us to measure the direction-resolved charge flow through this device \cite{Fujisawa06} and consequently the current probability distribution. A recent experiment following this line \cite{Utsumi10prb} revealed the importance of the backaction \cite{Gustavsson07a} of the charge sensing device, which was a quantum point contact \cite{Field93} (QPC). Backaction due to nonequilibrium QPC noise destroys microreversibility in the DQD and leads to an apparent temperature that significantly exceeds the bath temperature \cite{Utsumi10prb,Cuetera11,Golubev11}. This sensitivity to the measurement arises because the entropy fluctuations associated with single-electron tunneling are 3 orders of magnitude smaller than those at room temperature. In order to avoid the spurious backaction, we employ an optimized sample design combining electron-beam and scanning-probe lithography \cite{Roessler10}. It provides the high tunability and electronic stability required for the experiment while maintaining a good QPC--DQD coupling. We observe quantitative agreement between our data and theory in the near-equilibrium regime after including a small correction due to finite detector bandwidth \cite{Utsumi10book}, which is a technical rather than physical complication and requires no fitting parameter. In the regime far from equilibrium, our results depend on the system details; we find that the FT describes our data correctly in configurations where the DQD dynamics are those of a three-state Markovian system.

\begin{figure}[tb]
\includegraphics{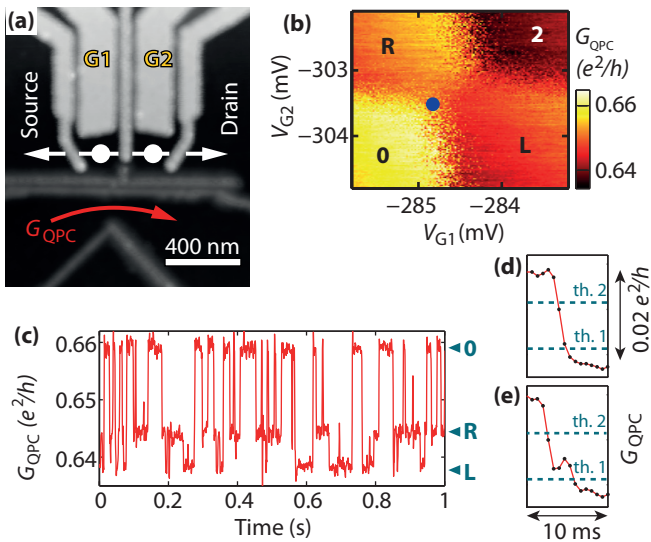}
\caption{(a) Atomic-force micrograph of the sample. Electrons can travel between the source and drain via the two quantum dots marked by disks. The conductance $G_\mathrm{QPC}$ of the quantum point contact serves to read out the charge state of the quantum dots. (b) By using the two gates G1 and G2, the charge state of the sample is controllably switched between empty (0), left (L), right (R) and doubly (2) occupied. These four states are visible as panels of distinct $G_\mathrm{QPC}$ in the plot. (c) Time dependence of $G_\mathrm{QPC}$ close to the charge degeneracy point marked by a dot in (b), displaying fluctuations between three levels: L, R, and 0. (d,e) $G_\mathrm{QPC}$ time segments showing examples of a direct transition from 0 to L (d), and of a transition from 0 to L via R (e).}
\label{fig:FlucThm_figure_Intro}
\end{figure}

Our measurements were performed in a $^3$He/$^4$He dilution refrigerator on the sample shown in Fig.~\ref{fig:FlucThm_figure_Intro}(a). The dark parts in the atomic-force micrograph correspond to the conductive (non-depleted) parts of a two-dimensional electron gas $34 \, \mathrm{nm}$ below the surface of a $\mathrm{GaAs/Al_{0.3}Ga_{0.7}As}$ heterostructure (sheet density $n_S = 4.9 \times 10^{15} \,\mathrm{m}^{-2}$, mobility $\mu = 33 \, \mathrm{m^2V^{-1}s^{-1}}$ as determined at $T = 4.2 \, \mathrm{K}$). Confinement is achieved in one part with Ti/Au gates (the upper half of the image) biased with negative voltages. The thin vertical finger gates are only slightly biased in order to maintain a small tunneling coupling between the two QDs (white disks) and the source and drain leads. The horizontal lines are created by local anodic oxidation; they electrically separate the DQD from the charge detector QPC in the lower half of the image. The detector's conductance $G_\mathrm{QPC}$ is sensitive to the charge on the DQD and abruptly decreases if an electron is loaded to either of the two QDs. Figure \ref{fig:FlucThm_figure_Intro}(b) shows a color plot of the time-averaged QPC conductance as a function of the two gate voltages $V_\mathrm{G1}$ and $V_\mathrm{G2}$, which control the electron number on the DQD \cite{Vanderwiel03}. Four regions of stable charge are visible as regions of constant QPC conductance, up to a background that is linear in $V_\mathrm{G1}$ and $V_\mathrm{G2}$. We estimate that each QD holds about 80 electrons, but for ease of notation we consider only the excess electron number and denote the four relevant charge states of the DQD as empty (0), singly occupied (L, R), and doubly occupied (2). At the borders of these regions, thermal fluctuations of the charge occur.

In particular, at the charge degeneracy point marked by a dot in the color plot, the QPC conductance rapidly fluctuates between three levels corresponding to the states 0, L, and R, as shown in Fig.~\ref{fig:FlucThm_figure_Intro}(c). By counting the number of transitions between L and R, we can determine the total number $n$ of electrons that pass through the center barrier of the DQD during an acquisition time $\tau$. If $\tau$ is large compared to the typical dwell time of an electron inside the DQD, the electron passing through the center barrier will typically reach one of the leads, where it equilibrates with the thermal bath at temperature $T$. The entropy change $\Delta S$ will then be equal for all charge fluctuations with equal $n$ \cite{Esposito07,Saito08,Esposito09}. The change can be either positive or negative, depending on the direction of the charge flow with respect to the bias direction. As the dissipated heat $neV_\mathrm{DQD}$ is determined by the DQD source-drain voltage, $\Delta S$ is given by $neV_\mathrm{DQD}/T$, and the FT for our system is
\begin{equation}
\label{eq:FlucThm_FT_DQD}
\frac{P_{\tau}(n)}{P_{\tau}(-n)} = e^{neV_\mathrm{DQD}/k_BT}.
\end{equation}

The QPC is biased with a voltage of $300 \, \mathrm{\mu V}$ and its conductance recorded using a room-temperature current-to-voltage converter and digitizer. The signal is then filtered by software at a bandwidth of $0.4 \, \mathrm{kHz}$, further resampled at $1.5 \, \mathrm{kHz}$, and stored for analysis. Two conductance thresholds were defined in the middle between neighboring conductance levels. In the algorithm to determine the chronological sequence of DQD states, we build in the requirement that at least three successive data points must lie in the same conductance interval for the DQD state to be accepted. The reason for this choice becomes clear when taking a closer look at the $G_\mathrm{QPC}$ time traces provided in Fig.~\ref{fig:FlucThm_figure_Intro}(d) and (e). The $G_\mathrm{QPC}$ signal in panel (d) exhibits a direct transition from level 0 to level L. Because of to the finite rise time, two of the sampled data points happen to lie between thresholds 1 and 2. If these were to be assigned to level R, a false transition from R to L would be counted, reducing the net flow $n$ in the corresponding time segment by 1. In comparison, the signal in panel (e) shows a short, yet clear, dwell time in the level R.

\begin{figure}[tb]
\includegraphics{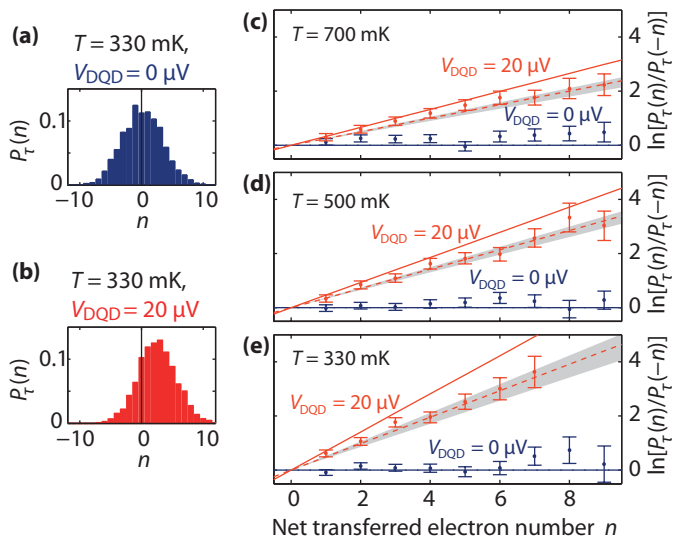}
\caption{(a,b) Experimental probability distributions for the net electron number $n$ transferred through the DQD during time $\tau = 2 \, \mathrm{s}$ for two different DQD source-drain voltages $V_\mathrm{DQD}$. (c,d,e) Comparison of experimental data with theory for three different bath temperatures. The data points correspond to the left-hand side of Eq.~\eqref{eq:FlucThm_FT_DQD} and describe the probability ratio of forward ($+n$, entropy-producing) and backward ($-n$, entropy-consuming) processes for a given $n$. The solid lines mark the expected exponential behavior $\exp (neV_\mathrm{DQD}/k_BT)$ for the two source-drain voltages $0 \, \mathrm{\mu V}$ (dark blue, horizontal) and $20 \, \mathrm{\mu V}$ (red, inclined). If the finite bandwidth of the detector is taken into account \cite{Naaman06, Utsumi10book} (dashed lines), experiment and theory agree within the statistical uncertainty of the data. (Error bars indicate the estimated standard deviation. The gray bands around the dashed lines indicate the uncertainty in the finite-bandwicth correction.)}
\label{fig:FlucThm_figure_TempDependence}
\end{figure}

Figure \ref{fig:FlucThm_figure_TempDependence}(a) shows an example of a $P_\tau(n)$ distribution measured at an electron-bath temperature of $330 \, \mathrm{mK}$ and with $V_\mathrm{DQD} = 0 \, \mathrm{\mu V}$. It is based on the counting analysis of 3000 $G_\mathrm{QPC}$ time segments, each with length $\tau = 2 \, \mathrm{s}$. The choice of $\tau$ is such to minimize the combined error that originates both from the imperfect long-time limit \cite{Esposito07,Saito08,Esposito09,Utsumi10prb} (favoring large $\tau$) and from statistics (favoring small $\tau$). The distribution is symmetric about $n=0$; i.e., there is no net charge flow, as expected in equilibrium. The amount of charge flow at zero DQD bias (compared to charge flow at finite DQD bias) is a sensitive measure for the strength of residual QPC backaction. Because of left-right asymmetries in the DQD such as a nonzero level detuning, QPC backaction generically leads to nonequilibrium charge flow in a ratchet-type effect \cite{Khrapai06,Gasser09}. 

When $V_\mathrm{DQD}$ is increased to $20 \, \mathrm{\mu V}$, the distribution shifts towards positive $n$, as shown in panel (b), and, on average, an electron number of $\langle n \rangle = 2.5$ is transferred. Still, for some of the time segments, the charge flow is against the applied bias ($n<0$), which results in a temporary decrease of the system entropy. Similar measurements were carried out at temperatures $500 \, \mathrm{mK}$ and $700 \, \mathrm{mK}$ \footnote{These electronic temperatures were determined from the width of thermally broadened Coulomb blockade resonances. The corresponding cryostat temperatures were $300$, $500$, and $700 \, \mathrm{mK}$, respectively. At even higher bath temperatures, both the increasing transition rates and the growing double occupancy of the DQD render the analysis difficult.}. These are shown in Fig.~\ref{fig:FlucThm_figure_TempDependence}(c--e), where the data points are the logarithm of the left-hand side of Eq.~\eqref{eq:FlucThm_FT_DQD}, measured at $V_\mathrm{DQD} = 0 \, \mathrm{\mu V}$ and $V_\mathrm{DQD} = 20 \, \mathrm{\mu V}$, respectively. The expression $\ln [P_\tau(n)/P_\tau(-n)]$ follows the expected linear behavior close to the theoretical curve $neV_\mathrm{DQD}/k_BT$ (solid lines).

In the nonzero-bias case, there is a systematic deviation of 20\% to 30\% in the slope. This can be understood by taking into account the limited bandwidth of the charge detection. A charge-switching event in the DQD is detected in the QPC only after a reaction time of $1/\Gamma_\mathrm{det}$, which in our case is determined both by the rise time of the measurement electronics and by the rejection of short events built into the analysis algorithm. If the charge state switches back too fast, the event is missed. Following the ideas presented in Ref.~\cite{Naaman06}, Utsumi \emph{et al.}~\cite{Utsumi10book} calculated the effect of the finite bandwidth. They found that, up to order $1/\Gamma_\mathrm{det}$, the finite bandwidth has the same effect as a prefactor $\alpha_\mathrm{BW}<1$ to the term $eV_{DQD}/k_BT$, just as we have observed in our experiment. The factor $\alpha_\mathrm{BW} = k_B T \ln w^{*}/eV_\mathrm{DQD}$ is expressed in terms of the six transition rates $\Gamma_{ij}$ between the states $i,j = \mathrm{L,\,R,\,0}$,
\begin{equation}
\label{eq:FlucThm_finiteBW}
w^{*} \approx w+\frac{1-w}{\Gamma_\mathrm{det}} \left( \frac{\Gamma_\mathrm{L0}\Gamma_\mathrm{0R}}{\Gamma_\mathrm{LR}} +
\frac{\Gamma_\mathrm{RL}\Gamma_\mathrm{L0}}{\Gamma_\mathrm{R0}} +
\frac{\Gamma_\mathrm{0R}\Gamma_\mathrm{RL}}{\Gamma_\mathrm{0L}}  \right) ,
\end{equation}
where $w=\exp(eV_\mathrm{DQD}/k_BT)$. Qualitatively, the effect can be understood as follows. At a nonzero bias, transitions directed toward the drain occur with faster rates ($\Gamma_\mathrm{L0}$, $\Gamma_\mathrm{RL}$, $\Gamma_\mathrm{0R}$) than those directed toward the source ($\Gamma_\mathrm{0L}$, $\Gamma_\mathrm{LR}$, $\Gamma_\mathrm{R0}$). It is therefore more likely that the detector misses a charge flowing toward the drain than toward the source. In short, the ratio $P_\tau(n)/P_\tau(-n)$ becomes underestimated for $n>0$. It is important to stress that, despite the similar phenomenology, the correction we employ here has a very different quality than the effective temperature due to QPC backaction introduced in Ref.~\cite{Utsumi10prb}. QPC backaction implies excitation of degrees of freedom in the microscopic system of interest, whereas the effect of finite bandwidth is a matter of the imperfect room-temperature electronics and leaves the microscopic system unperturbed.

\begin{figure}[tb]
\includegraphics{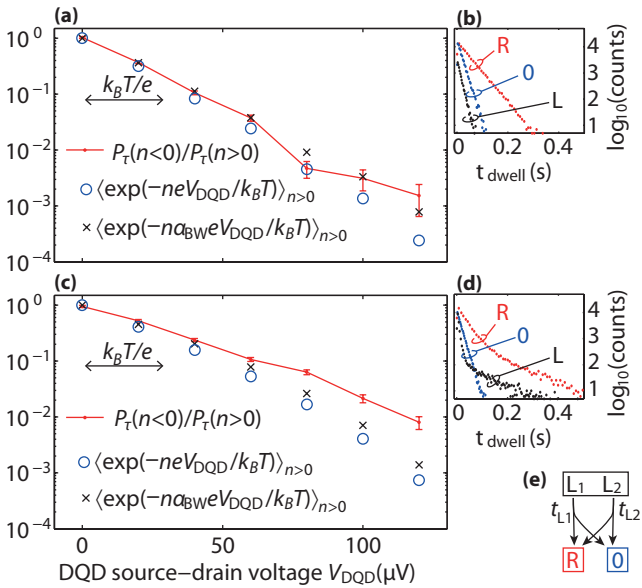}
\caption{(a) The red data points show the $V_\mathrm{DQD}$ depencence of the left-hand side of Eq.~\eqref{eq:FlucThm_Integrated}, which is the ratio of entropy-consuming vs.~entropy-producing cycles, with error bars that indicate its estimated standard deviation. The blue circles show the right-hand side, which is the average of the Boltzmann factor among the entropy-consuming cycles. The FT [Eq.~\eqref{eq:FlucThm_Integrated}] is satisfied if the finite detector bandwidth is taken into account in the form of a correction factor $\alpha_\mathrm{BW}$ to the exponent in Eq.~\eqref{eq:FlucThm_Integrated} (shown as crosses). The uncertainty in the finite-bandwidth correction is comparable with the error in $P_\tau(n<0)/P_\tau(n>0)$ for all bias voltages. (b) Dwell-time distributions of the $G_\mathrm{QPC}$ signal in the states L, R, and 0, as extracted from the data set in (a) at the point $V_\mathrm{DQD} = 0 \, \mathrm{\mu V}$. (c,d) Same as (a) and (b), but measured at a slightly different gate-voltage configuration. In this case, the dwell times at zero bias (d) are nonexponentially distributed for the states L and R. The finite-bandwidth model [the crosses in (c)] is not valid for this case but plotted for comparison. (e) Diagram of the decay of a charge state L (black rectangle) that consists of two internal QD states L1 and L2. Such a process can lead to nonexponentially distributed charge dwell times as shown in (d).}
\label{fig:FlucThm_figure_BiasDependence}
\end{figure}

In the finite-bandwidth model, $\Gamma_\mathrm{det}$ plays the role of a mean inverse reaction time of the detector. We use a simulation to determine this parameter for our particular detection scheme. A square pulse of duration $\tau_p$ is filtered and resampled the same way as the experimental signal. Our value of $\Gamma_\mathrm{det} = (0.59 \pm 0.12) \, \mathrm{kHz}$ is then defined as the inverse of the minimum $\tau_p$ for which at least three sampled points reach 50 \% of the square-pulse amplitude. The dashed lines in Figs.~\ref{fig:FlucThm_figure_TempDependence}(c--e) show the theoretical expectation calculated with this value for $\Gamma_\mathrm{det}$ and the experimentally determined  $\Gamma_{ij}$ \cite{Fujisawa06}, and indeed agree much better with the experiment. The gray shaded areas indicate the uncertainty in the slope which is mainly determined by the uncertainty in $\Gamma_\mathrm{det}$. We emphasize that this analysis does not involve any free parameters.

The DQD voltage of $20 \, \mathrm{\mu V}$ used in the temperature-dependence measurements is comparable with the thermal voltage $k_B T/e = 28...58 \, \mathrm{\mu V}$, so the system is not too far from thermal equilibrium. The FT also applies far away from equilibrium, however. To test its predictions in this regime, we have also performed bias-dependence measurements with $V_\mathrm{DQD}$ up to $120 \, \mathrm{\mu V}$, i.e., about $4.3 \times k_B T/e$ at $T = 330 \, \mathrm{mK}$. The data are shown in Fig.~\ref{fig:FlucThm_figure_BiasDependence} and are based on the analysis of 2000 $G_\mathrm{QPC}$ time segments of length $\tau = 2 \, \mathrm{s}$ for each DQD voltage. For clarity and to reduce the statistical error for large voltages, we plot an integrated version of the FT relating the total fractions of entropy-producing and entropy-consuming cycles,
\begin{equation}
\label{eq:FlucThm_Integrated}
\frac{\sum_{n<0} P_\tau (n)}{\sum_{n>0} P_\tau (n)} = \frac{\sum_{n>0} P_\tau (n) e^{-neV_\mathrm{DQD}/k_BT}}{\sum_{n>0} P_\tau (n)}.
\end{equation}

In our measurements of the $V_\mathrm{DQD}$ dependence, we keep the gate voltages $V_\mathrm{G1}$ and $V_\mathrm{G2}$ fixed. The choice of $V_\mathrm{G1}$ and $V_\mathrm{G2}$ determines the level arrangement of the DQD with respect to the electrochemical potentials of the leads and is \emph{a priori} not relevant for the validity of Eq.~\eqref{eq:FlucThm_Integrated}. In our measurements we observe a rather strong effect of this choice which is not fully understood. Thus, we plot in Figs.~\ref{fig:FlucThm_figure_BiasDependence}(a) and (c) the data of two representative measurements. The differing gate-voltage configurations in the two measurements result in differing relative arrangements of the electrochemical potentials of source, drain, left, and right QD (denoted by $\mu_S=eV_\mathrm{DQD}/2$, $\mu_D = -eV_\mathrm{DQD}/2$, $\mu_L$, and $\mu_R$, respectively). In measurement (a), we have $\mu_L-\mu_R \approx 80 \, \mathrm{\mu eV}$ and $(\mu_L+\mu_R)/2 \approx 10 \, \mathrm{\mu eV}$, whereas in measurement (b), we have $\mu_L-\mu_R \approx 55 \, \mathrm{\mu eV}$ and $(\mu_L+\mu_R)/2 \approx -5 \, \mathrm{\mu eV}$, as determined by finite-bias spectroscopy.

The red data points in Fig.~\ref{fig:FlucThm_figure_BiasDependence}(a) plot the left-hand side of Eq.~\eqref{eq:FlucThm_Integrated}. The quantity rapidly decreases with the voltage, as charge transfer against the bias occurs less and less frequently. Measurements at even higher DQD voltages are eventually limited by the necessary, exponentially increasing measurement time. The blue circles plot the right-hand side, calculated without the finite-bandwidth correction. Similar to the low-bias case, there is a systematic deviation. For the black crosses, the exponent in Eq.~\eqref{eq:FlucThm_Integrated} is replaced with the bandwidth-corrected version $\alpha_\mathrm{BW} neV_\mathrm{DQD}/k_BT$, and we see that the observed deviation can entirely be attributed to this measurement issue.

In the second measurement shown in Fig.~\ref{fig:FlucThm_figure_BiasDependence}(c), there is a larger discrepancy between the two sides of Eq.~\eqref{eq:FlucThm_Integrated} which goes beyond the effect of the finite detector bandwidth. We observe that this discrepancy coincides with the presence of nonexponential distributions of the random dwell times in the three DQD charge states. Figures \ref{fig:FlucThm_figure_BiasDependence}(b) and (d) show the histograms of the signal dwell times in L, R, and 0 measured at $V_\mathrm{DQD} = 0 \, \mathrm{\mu V}$ for the two configurations. In configuration (a), the dwell times follow exponential distributions with a single lifetime. This is not the case in configuration (c), where the states L and R are not characterized by a single lifetime. This is indicative for the population of additional (excited) states in the DQD \cite{Gustavsson06b}. Considering the example of the left QD, a nonzero population $p_\mathrm{L2}$ of a long-lived excited state L2 means that the charge state L can no longer be identified with the QD ground state L1. The QPC, which monitors the dynamics of the charge states, sees a dwell time $t$ in the state L distributed according to $p_{L1}t_\mathrm{L1}^{-1}\exp(-t/t_\mathrm{L1})+p_{L2}t_\mathrm{L2}^{-1}\exp(-t/t_\mathrm{L2})$, where $t_\mathrm{L1}$ and $t_\mathrm{L2}$ are the decay times of the two QD states into R and 0 [cf.~Fig.~\ref{fig:FlucThm_figure_BiasDependence}(e)]. In configuration (c), the QD energies are lower than in configuration (a), which makes population of excited states more probable. Although the precise level arrangement cannot be reconstructed from our data, the measurement in Fig.~\ref{fig:FlucThm_figure_BiasDependence}(c) demonstrates the sensitivity of such a test of the FT to the details of the DQD level structure.

In conclusion, we have presented an extensive quantitative test of the fluctuation theorem for electron transport through a DQD, covering different temperatures and strong nonequilibrium transport conditions. Our results validate the theory in the near-equilibrium regime with a good accuracy. A remaining discrepancy is very well explained with a master-equation model of the finite-bandwidth detection. This agreement proves the usefulness of this corrective approach in compensating for a slow detector. In the regime far from equilibrium, our results display a strong dependence on the internal DQD level structure controlled by gate voltages. In configurations where our system is well described as a three-state Markovian system, we observe a good agreement with theory, demonstrating the potential of the DQD as a model system for the study of nonequilibrium thermodynamics. Our results anticipate the test of the FT in quantum-coherent electron transport, which requires the measurement of thermal fluctuations on a sub-Kelvin energy scale.

The authors thank K.~Kobayashi, A.~Yacoby, C.~Flindt, and G.~Sch\"on for discussions. Sample growth and processing was mainly carried out at FIRST laboratory, ETH Zurich. Financial support from the Swiss National Science Foundation (Schweizerischer Nationalfonds) is gratefully acknowledged.

\end{document}